# Spectral Efficiency Optimization for an Interfering Cognitive Radio with Adaptive Modulation and Coding

Mehrdad Taki and Farshad Lahouti
Wireless Multimedia Communications Laboratory
School of Electrical and Computer Engineering, University of Tehran, Iran
[mehrdadtaki, lahouti]@ut.ac.ir, http://wmc.ut.ac.ir

*Abstract*— **In this paper, we consider a primary and a cognitive user transmitting over a wireless fading interference channel. The primary user transmits with a constant power and utilizes an adaptive modulation and coding (AMC) scheme satisfying a bit error rate requirement. We propose a link adaptation scheme to maximize the average spectral efficiency of the cognitive radio, while a minimum required spectral efficiency for the primary user is provisioned. The resulting problem is constrained to also satisfy a bit error rate requirement and a power constraint for the cognitive link. The AMC mode selection and power control at the cognitive transmitter is optimized based on the modified signal to noise plus interference ratio feedback of both links. The problem is then cast as a nonlinear discrete optimization problem for which a fast and efficient suboptimum solution is presented. We also present a scheme with rate adaptive and constant power cognitive radio. An important characteristic of the proposed schemes is that no computation or coordination overhead is imposed on the primary radio due to the cognitive radio activity. Numerical results and comparison with the interweave approach to cognitive radio demonstrate the efficiency of the proposed solutions.**

## I. INTRODUCTION

Cognitive radio, as a promising technology to improve spectrum utilization efficiency, has recently been the subject of intensive research. In this method, a secondary (cognitive) link is activated along with the primary (licensed) link in a way that it does not disrupt the primary link. There are three well-known approaches to the cognitive transmission, namely the interweave, the underlay, and the overlay approaches [1]. In the interweave approach, the cognitive user identifies spectrum gaps that are not in use by the licensed users for transmission. In the underlay approach, the cognitive radio is aware of the gain of its channel to the primary receiver and transmits in a manner that the resulting interference is negligible. In the overlay approach, the cognitive radio imposes non-negligible interference on the primary receiver but it makes up the performance degradation at the primary radio with the aid of its non-causal access to the primary user data [2].

In information theory, it is shown that in an interference channel with fixed link gains, higher rates can be achieved by interfering transmission and subsequently, canceling part of the interference at the receivers [3]. In general, this is accomplished by use of coordinated coding schemes at the transmitters. In typical cognitive radio applications, however, this may not be feasible as the primary user is not necessarily aware of the cognitive user activity. In addition, such approach to interference cancelation incurs increased complexity and security concerns, as it involves decoding of unintended signal at the receivers. In this case, treating interference as noise is a simple solution that is known to outperform orthogonal transmission in a weak interference channel [4]. Moreover, it is shown that this approach achieves the system sum capacity, when the ratio of interference to desired signal power is smaller than a threshold (noisy interference) [5].

Rate adaptation is an enabling technology for improved transmission performance over wireless fading links and is already included in several recent wireless communication standards [6]. The advantage of rate adaptation in cognitive radio is analyzed for several schemes based on continuous rate transmission in [7]-[9]. In continuous rate transmission, the rate is continuously controlled by varying the transmission power. In practical rate adaptive systems, a limited number of transmission rates are set up based on different pairs of modulation and channel coding schemes. This is known as adaptive modulation and coding (AMC) and may also be used in conjunction with power control (AMCP) for improved performance [10]. The AMCP is employed in [11] to improve the spectral efficiency of a cognitive radio network based on OFDMA in an interweave setting.

In this paper, we propose a new scheme for the interfering cognitive radio transmission in a wireless fading system, where the interference is treated as noise at the unintended receiver. The proposed scheme utilizes AMCP to maximize the average spectral efficiency of the cognitive link, satisfying a minimum required average spectral efficiency for the corresponding primary link. The power constraint on the cognitive transmitter and the bit error rate (BER) requirement for the cognitive and primary links are provisioned. This leads to a nonlinear mixed-integer optimization problem. To solve this, we reformulate it into an integer nonlinear optimization problem for which an efficient solution is proposed. The optimization procedure is performed at the cognitive radio based on the signal to noise plus interference ratios (SNIR) of both links. The SNIR of the cognitive link is estimated at the corresponding receiver. The SNIR of the primary link may be estimated by the cognitive radio, e.g., by eavesdropping on the primary link feedback [12]. However, there is no need to extra computations at the primary radio for extracting the gain of cross link from the cognitive transmitter to the primary receiver and communicate it to the cognitive radio [13]. In this paper, we also present a constant power adaptive rate interfering scheme, which operates based on even smaller negotiation overhead. Numerical results and comparisons with the interweave approach to the cognitive radio are presented, which demonstrate the efficiency of the proposed schemes.

The rest of this article is organized as follows. Section II provides the system model, adaptive transmission strategy, and the BER performance approximation for the links. Section III describes the proposed interfering cognitive transmission schemes. In section IV, the interweave approach to cognitive transmission employing AMCP is analyzed. Section V presents numerical results and comparisons. We conclude this article in Section VI.

## II. PRELIMINARIES

### A. Notations

In this paper, lower case italic letters denote random variables, e.g. $z$. Functions are denoted by lower case letters, e.g., $g(.)$, constant parameters are shown by uppercase letters, e.g., N and sets are denoted by scripts, e.g. $\mathcal{C}$. The probability density function (PDF) of a random variable $x$ is denoted by $f_x(.)$. The probability of the event $A$ is shown by $\text{prob}\{A\}$.

### B. System Description and Channel Model

Fig. 1 depicts the wireless network system model under consideration. There are two interfering wireless links, namely the primary and cognitive links, indexed as 1 and 2, respectively. Each link involves a user with a transmitter $\text{Tx}_i$ wishing to communicate with a corresponding receiver $\text{Rx}_i$, $i = 1,2$. The channels are assumed to be discrete-time memoryless, such that the received signals depend on the transmitted signals as follows:

$$y_i = h_{ii}x_i + h_{ji}x_j + \eta_i; \quad i,j = 1,2; \quad i \neq j \tag{1}$$

where $x_i$ is the transmitted signal from $\text{Tx}_i$ and $\eta_i$ denotes the additive white Gaussian noise at $\text{Rx}_i$. The terms $h_{ij}$ denote the gains of direct ($i = j$) and cross links ($i \neq j$). The interference that is imposed by the unintended transmitter is given by $h_{ji}x_j$. We assume frequency flat fading channels with stationary and ergodic time-varying gains. A block-fading model is adopted, where the channel gain remains constant during a block-length (here a codeword), and independently changes from one block to another [15].

## C. Bit Error Rate Approximation

As discussed in section I, due to complexity, security and other practical considerations, in this work we assume that the interference of the unintended transmitter is treated as noise at each receiver. The SNIR at $Rx_i$ is given by:

$$\gamma_i = \frac{p_i s_{ii}}{p_j s_{ji} + N_0} \qquad i,j = 1,2; i \neq j \tag{2}$$

where $N_0$ is the variance of AWGN, $p_i$ is the transmission power of $Tx_i$ and $s_{ji} = |h_{ji}|^2$. In an AMC system, there are $N + 1$ transmission modes, each characterized by a modulation and a coding rate, resulting in a transmission rate, $R_n$ [10]. The AMC modes are assumed to be sorted according to their rates, i.e.,

$$0 = R_0 < R_1 < R_2 \ldots < R_N. \tag{3}$$

The mode "0" corresponds to no data transmission or outage. The BER performance of the signaling when the link SNIR is $\gamma$, may be approximated by a fitting expression as follows [16]:

$$p_e(\gamma, R_n) = A_n \cdot \exp(-A'_n \times \gamma), \quad 0 \leq \gamma, \tag{4}$$

where $\{A_n, A'_n\}$ are mode specific constants. In transmission mode $n$, the minimum required SNIR to guarantee an instantaneous BER smaller than a predetermined value $B_0$, is given by $g_{B_0}(R_n)$. We have:

$$p_e(\gamma, R_n) \leq B_0 \Longrightarrow \gamma \geq g_{B_0}(R_n), \tag{5}$$

$$g_{B_0}(R_n) \stackrel{\text{def}}{=} -1/A'_n \times \ln(B_0/A_n), \qquad B_0 \leq A'_n. \tag{6}$$

As elaborated in the followings, the AMC mode selection and power control for the primary and cognitive users are accomplished based on the estimates of their direct link SNIRs denoted by $\hat{\gamma}_1$ and $\hat{\gamma}_2$, respectively. We assume that these estimates are obtained perfectly without delay.

## III. ADAPTIVE COGNITIVE TRANSMISSION SCHEMES

In this section, we propose adaptive transmission schemes for the presented system model to maximize the average spectral efficiency of the cognitive link, while satisfying a minimum required average spectral efficiency for the primary link and the power constraint on the cognitive transmitter. It is assumed that the application requires a maximum BER of $B_1$ for the primary link and $B_2$ for the cognitive link. The primary user transmits with a constant power $p_1 = P_1$. Without loss of generality, we assume that both users employ similar AMC transmission modes, i.e., sets of modulation and coding schemes. We propose two link adaptation schemes for cognitive radio in an interfering scenario; the first one assumes a constant power AMC and the second one employs adaptive power transmission (AMCP).

*A. Constant Power Link Adaptation Scheme*

In this section, we consider the scenario, where the cognitive user transmits with a constant power $p_2 = P_2$. Each of the two radios adapts its AMC rate based on its own link SNIR to satisfy its BER requirement. The assigned AMC rates for the primary and the cognitive links, based on the estimated SNIRs of the links, are set as $k_1(\hat{\gamma}_1)$ and $k_2(\hat{\gamma}_2)$, respectively. If the average power of the cognitive transmitter is limited to $P_2^{max}$, the link adaptation problem at the cognitive radio is formulated as:

$$\max_{P_2} \int_0^\infty k_2(\hat{\gamma}_2) \cdot f_{\hat{\gamma}_2}(\hat{\gamma}_2) \, d\hat{\gamma}_2 \qquad \text{subject to:} \qquad (7)$$

$$\begin{cases} C1: \int_0^\infty k_1(\hat{\gamma}_1) \cdot f_{\hat{\gamma}_1}(\hat{\gamma}_1) \, d\hat{\gamma}_1 \geq E_1 \\ C2: P_2 \leq PW_2 \\ C3: p_e(\gamma_1, k_1(\hat{\gamma}_1)) \leq B_1 \\ C4: p_e(\gamma_2, k_2(\hat{\gamma}_2)) \leq B_2 \end{cases}$$

where $\gamma_1$ and $\gamma_2$ are the link SNIRs in the next transmission block. For the link adaptation it is assumed that as in [10], the range of SNIR of the transmitter $i \in \{1,2\}$ is divided into $N + 1$ non-overlapping consecutive intervals, where interval $n$ is denoted by $[v_{i,n}, v_{i,n+1})$ for $0 \leq n \leq N$ and $v_{i,0} = 0, v_{i,N+1} = \infty$. If the SNIR falls in the interval $n$, the AMC transmission mode $n$ with rate $R_n$ is selected. The average spectral efficiency of each link is:

$$k_i^{avg} = \sum_{n=1}^{N} R_n \times \text{prob}\{v_{i,n} \leq \hat{\gamma}_i < v_{i,n+1}\}, i = 1,2 \qquad (8)$$

Aiming at maximum link utilization, considering a given average power and satisfying the BER constraint, we have:

$$v_{i,n} = \min_{\hat{\gamma}_i} \hat{\gamma}_i \quad \text{s.t.:} \quad \hat{\gamma}_i \geq g_{B_i}(R_n), \ i = 1,2 \implies v_{i,n} = g_{B_i}(R_n) \qquad (9)$$

Using (8) and (9), the optimization problem in (7) is restated as follows:

$$\max_{P_2} k_2^{avg} \quad \text{s.t.}: \begin{cases} \text{C1}: k_1^{avg} \geq E_1 \\ \text{C2}: P_2 \leq P_2^{max} \end{cases} \quad (10)$$

The BER constraints (C3 and C4) in (7) are now considered in the rate assignments. Note that increasing $P_2$, increases $k_2^{avg}$ and decreases $k_1^{avg}$. The solution for (10) is obtained by increasing $P_2$ from zero up to a point where either C1 or C2 is satisfied with equality. For rate adaptation at the cognitive radio and solving the associated optimization problem, it is necessary to obtain $k_1^{avg}$ and the cognitive link SNIR. The cognitive receiver can estimate its own link SNIR. It can also compute $k_1^{avg}$, e.g., by eavesdropping on the primary link feedback as suggested in [12]. Therefore, the cognitive receiver solves the optimization problem and simply feeds back the desired AMC mode index.

*B. Variable Power Link Adaptation Scheme*

In the second scheme, the cognitive radio adapts its power $p_2 = p_2(\hat{\gamma}_1, \hat{\gamma}_2)$ and rate $k_2(\hat{\gamma}_1, \hat{\gamma}_2)$ to maximize the average spectral efficiency of its link, satisfying a minimum required average spectral efficiency for the primary link, the power constraint on the cognitive transmitter and the BER constraints on both links. The primary radio adapts its rate $k_1(\hat{\gamma}_1)$ independently and with the constant power $P_1$. In the followings, we first set up the link adaptation problem and next reformulate it in a form for which an effective solution is presented. The solution sets the values of the variables $k_2(.)$, $p_2(.)$ and $k_1(.)$ as functions of $\hat{\gamma}_1$ and $\hat{\gamma}_2$.

$$\max_{k_2(.)} \int_0^\infty \int_0^\infty k_2(\hat{\gamma}_1, \hat{\gamma}_2) f_{\hat{\gamma}_1,\hat{\gamma}_2}(\hat{\gamma}_1, \hat{\gamma}_2) d\hat{\gamma}_1 d\hat{\gamma}_2 \quad \text{s.t.}: \quad (11)$$

$$\begin{cases} \text{C1}: \int_0^\infty k_1(\hat{\gamma}_1) f_{\hat{\gamma}_1}(\hat{\gamma}_1) d\hat{\gamma}_1 \geq E_1 \\ \text{C2}: \int_0^\infty \int_0^\infty p_2(\hat{\gamma}_1, \hat{\gamma}_2) f_{\hat{\gamma}_1,\hat{\gamma}_2}(\hat{\gamma}_1, \hat{\gamma}_2) d\hat{\gamma}_1 d\hat{\gamma}_2 \leq PW_2 \\ \text{C3}: p_e(\gamma_1, k_1(\hat{\gamma}_1)) \leq B_1 \\ \text{C4}: p_e(\gamma_2, k_2(\hat{\gamma}_1, \hat{\gamma}_2)) \leq B_2 \end{cases}$$

The power adapted SNIRs of links in the current transmission are denoted by $\gamma_1$ and $\gamma_2$. For rate adaptation at the cognitive radio, the optimization problem (11) is to be solved, which requires the knowledge of $\hat{\gamma}_1$ and $\hat{\gamma}_2$. The cognitive receiver can estimate its own link SNIR. It can also compute $\hat{\gamma}_1$, e.g., by eavesdropping [12] on the primary link feedback of SNIR or mode index. However, as evident, (11) is a complicated nonlinear mixed-integer optimization problem and cannot be directly solved.

*1) Problem Formulation*

In this section, we reformulate the desired optimization problem in (11) in a more tractable form. This is done by re-examining the SNIRs and exploiting the discrete nature of AMC transmission rates.

First, we assume that the additive thermal noise at the receivers is in general negligible, when its power is compared to that of the interference signal. This assumption during the design stage may result in a violation of the constraints C3 and C4 in (11). To address this issue, one may consider a tighter BER constraint than that required by the application. The received SNIRs is then given by

$$\begin{cases} p_2(.) > 0 \Rightarrow \begin{cases} \gamma_1 \cong P_1 s_{11}/(p_2(.)s_{21}) = \alpha\, P_1/p_2(.) \\ \gamma_2 \cong p_2(.)s_{22}/(P_1 s_{12}) = \beta p_2(.)/P_1 \end{cases} \\ p_2(.) = 0 \Rightarrow \begin{cases} \gamma_1 = P_1 s_{22}/N_0 \\ \gamma_2 = 0 \end{cases} \end{cases} \quad (12)$$

Considering the above equation, the modified SNIRs of the primary link ($\alpha \overset{\text{def}}{=} s_{11}/s_{21}$) and the cognitive link ($\beta \overset{\text{def}}{=} s_{22}/s_{12}$) may be estimated for the subsequent transmission. Subsequently, the rate of the cognitive radio is assigned based on them as $k_2(\alpha, \beta)$.

Satisfying C4 in (11) with equality, noting Eq. (12), the power of the cognitive transmitter is given by

$$p_2(\alpha, \beta, k_2(\alpha, \beta)) = P_1\, g_{B_2}(k_2(\alpha, \beta))/\beta, \quad k_2(.) > 0. \quad (13)$$

This transmission power results in the following SNIR at the primary receiver

$$\gamma_1 = \alpha\beta/g_{B_2}(k_2(\alpha, \beta)). \quad (14)$$

The primary user selects the maximum rate from AMC table that satisfies its BER requirement, i.e.,

$$k_1(\alpha, \beta, k_2(\alpha, \beta)) = \arg\max_{R_n} g_{B_1}(R_n) \quad \text{s.t.}: \; g_{B_1}(R_n) \leq \gamma_1, k_2(.) > 0. \quad (15)$$

Note that based on Eq. (14), $\gamma_1$ (similarly its estimate $\hat{\gamma}_1$) is a function of $\alpha, \beta$ and $k_2(\alpha, \beta)$, therefore, $k_1(\hat{\gamma}_1)$ is alternatively represented as in LHS of (15). Considering (12), (14) and (15), we then have

$$g_{B_1}(k_1(.)) \times g_{B_2}(k_2(.)) \leq \alpha\beta = \gamma_1\gamma_2 \quad, k_2(.) > 0. \quad (16)$$

The optimization problem in (11) is now restated as follows:

$$\max_{k_2(\alpha,\beta)} \int_0^\infty \int_0^\infty k_2(\alpha,\beta) f_\alpha(\alpha) f_\beta(\beta)\, d\alpha d\beta \quad \text{s.t.}:$$

$$\begin{cases} C1: \int_0^\infty \int_0^\infty k_1(\alpha,\beta,k_2(\alpha,\beta)) f_\alpha(\alpha) f_\beta(\beta)\, d\alpha d\beta \geq E_1 \\ C2: \int_0^\infty \int_0^\infty P_1\, g_{B_2}(k_2(\alpha,\beta))/\beta \times f_\alpha(\alpha) f_\beta(\beta)\, d\alpha d\beta \leq P_2^{max} \end{cases} \quad (17)$$

where the BER constraints, C3 and C4, in (11) are now considered in the power and rate assignments as in

Eqs. (13) and (15). It is clear from the definitions that $\alpha$ and $\beta$ are independent random variables. From Lemma 1 presented in Appendix A, the PDFs of the modified SNIRs for the system in Fig. 1 with Rayleigh fading channels may be obtained.

*Remark:* When $k_2(.) = 0$ and hence $p_2(.) = 0$, the primary user selects its rate based on its link SNR. Therefore, its average rate, considering the BER constraint, is given by

$$k_1(\alpha, \beta, k_2(\alpha, \beta) = 0) = \sum_{n=1}^{N} R_n \times \text{prob}\{v_{1,n} \leq P_1 s_{11}/N_0 < v_{1,n+1} | \alpha, \beta\}, \tag{18}$$

where $v_{1,n}$ is obtained in (9).

*Remark:* Noting Eqs. (12) and (15), when the cognitive radio transmission power is $p_2(.) > P_1 \alpha / g_{B_2}(R_1)$, the primary link BER requirement, even with its lowest AMC rate, is violated and hence an outage occurs.

We next consider the discrete nature of the AMC scheme to convert the problem to a discrete and manageable form. To this end, we consider several definitions.

*Definition: Permissible rate pairs set (rate set)*

Considering Eq. (16), there are certain rate pairs that are valid for the primary and cognitive links at each point in the $\alpha - \beta$ plane. This set of rate pairs is referred to as the permissible rate pairs set or simply the rate set of that point. The rate set also contains the case in which $k_2(.) = 0$.

*Definition: Common rate set regions*

As the primary and cognitive rates in Eq. (15) are discrete variables, certain regions partitioning the $\alpha - \beta$ plane are formed, in each the corresponding rate set remains constant. In general, there are $M = N^2 + 1$ such regions that are referred to as the common rate set regions. The common rate set region $i$ is defined as:

$$Z_{i-1} \leq \alpha\beta \leq Z_i, \quad 1 \leq i \leq M \tag{19}$$

where $Z_0 = 0$, $Z_M = \infty$ and $Z_i \in \{g_{B_1}(R_n) \times g_{B_2}(R_m), 1 \leq m, n \leq N\}$, for $1 \leq i \leq M-1$ and $Z_i$'s are sorted in ascending order $Z_0 < Z_1 \leq Z_2 \leq \cdots \leq Z_M$. In Fig. 2, for a given set of AMC modes, described in section V, the common rate set regions are identified as the area between two consecutive solid lines. These boundary lines correspond to $\alpha\beta = Z_i, 0 \leq i \leq M$.

*Definition: Common rate regions*

The common rate regions are areas in the $\alpha - \beta$ plane, each of which belongs to one common rate set region and is assigned the same rate pair. These are analogous to the non-overlapping intervals partitioning the SNR range in the single link AMC design. These areas are indentified by the boundary lines described above and sufficient auxiliary boundary lines ($L + 1$ radial lines and $C - M$ curves) as follows:

$$\beta/\alpha = W_j, \quad 0 \leq j \leq L \tag{20}$$

$$\alpha\beta = Z'_j \quad 1 \leq j \leq C - M \tag{21}$$

where $W_j$'s and $Z'_j$'s are constants and $W_0 = 0$, $W_L = \infty$. These lines are depicted as dashed lines in Fig. 2. Using the mentioned boundaries the $\alpha - \beta$ plane is divided into $V_0 = L \times C$ common rate regions that are denoted by $Reg(i), 1 \leq i \leq V_0$, and are simply called regions in the rest of the paper. As the rates in a given region are fixed, when the modified SNIRs fall into $Reg(i)$ (a common rate region), the rates assigned to the cognitive and primary links (two dependent variables) are stated as a function of region index and denoted by $k_2(i)$ and $k_1(i, k_2(i))$. Therefore, the average spectral efficiency of the primary and cognitive links, $k_1^{avg}$ and $k_2^{avg}$, are computed as follows:

$$k_2^{avg} = \sum_{i=1}^{V_0} k_2(i) \, pr(i) \tag{22}$$

$$k_1^{avg} = \sum_{i=1}^{V_0} k_1(i, k_2(i)) \, pr(i) \tag{23}$$

where $pr(i)$ is the probability that the modified SNIRs fall into $Reg(i)$.

*Remark:* If the cognitive link is inactive $k_2(i) = 0$, based on Eq. (14), the average rate of the primary link when modified SNIRs fall in $Reg(i)$ is:

$$k_1(i, k_2(i) = 0) = \sum_{n=1}^{N} R_n \times \text{prob}\{v_{1,n} \leq P_1 s_{11}/N_0 < v_{1,n+1} | \alpha, \beta \in Reg(i)\}. \tag{24}$$

Noting Eq. (13), if the normalized average power of the cognitive radio in $Reg(i)$, is defined as

$$p(i) \stackrel{\text{def}}{=} 1/pr(i) \times \iint_{Reg(i)} 1/\beta \times f_\alpha(\alpha) f_\beta(\beta) \, d\alpha \, d\beta, \tag{25}$$

then the average power that is used in the cognitive transmitter is given by

$$p_2^{avg} = P_1 \times \sum_{i=1}^{V_0} g_{B_2}(k_2(i)) \, p(i) \, pr(i). \tag{26}$$

In general, computation of $p(i)$, $pr(i)$ and $k_1(i, k_2(i) = 0)$ is complicated. In Appendix B, for the case of Rayleigh fading channels, these parameters are derived and simplified.

It is clear that when the number of regions goes to infinity, the summations in equations (22), (23) and (26) approach their corresponding values in (17). The desired optimization problem can now be restated as

follows:

$$\max_{k_2(i), 1 \leq i \leq V_0} k_2^{avg} = \sum_{i=1}^{V_0} k_2(i)\, pr(i) \quad \text{s.t.:} \tag{27}$$

$$\begin{cases} \text{C1: } k_1^{avg} = \sum_{i=1}^{V_0} k_1(i, k_2(i))\, pr(i) \geq E_1 \\ \text{C2: } p_2^{avg} = P_1 \sum_{i=1}^{V_0} g_{B_2}(k_2(i))\, p(i)\, pr(i) \leq P_2^{max} \end{cases}$$

*2) Problem Solution*

In general, one may consider solving the problem in (27) by integer nonlinear programming methods [17]. The optimality and convergence of these iterative algorithms rely on the proper definition of the gradient function and the second-order differentiation of the corresponding continuous form of the objective function. However, the discrete function, $k_1(i, k_2(i))$, does not have a closed form expression, yet a continuous form that is differentiable. In what follows, we propose a fast and simple alternative iterative algorithm that is inspired by gradient methods and enabled with a proper definition of the gradient function and selection of the initial point. To describe the proposed algorithm, we first present several definitions:

*Definitions: Decision Variables*

For each region, $1 \leq i \leq V_0$, if $k_2(i) = R_n > 0$, three decision variables are defined as follows. For $k_2(i) = 0$, these variables are set to zero.

$$d_1(i) \stackrel{\text{def}}{=} \begin{cases} 0 & , k_1(i, R_n) = R_N \\ -\dfrac{\Delta k_1^{avg}}{\Delta k_2^{avg}} = \dfrac{(k_1(i, R_n) - k_1(i, R_{n-t}))\, pr(i)}{(R_n - R_{n-t})\, pr(i)}, & k_1(i, R_n) < R_N \end{cases} \tag{28}$$

$$d_2(i, x) \stackrel{\text{def}}{=} \dfrac{\Delta p_2^{avg}}{\Delta k_2^{avg}} = \dfrac{(g_{B_2}(R_n) - g_{B_2}(R_{n-x}))\, p(i)\, pr(i)}{(R_n - R_{n-x})\, pr(i)} \tag{29}$$

$$d_3(i) \stackrel{\text{def}}{=} \begin{cases} 0, & k_1(i, R_n) = R_N \\ d_2(i, t), & k_1(i, R_n) < R_N \end{cases} \tag{30}$$

where $t$ in (28) and (30) is the minimum possible decrement in the index of the assigned rate to cognitive link that increases the rate of primary link, i.e.,

$$t = \min x \quad \text{subject to: } (k_1(i, R_{n-x}) - k_1(i, R_n)) > 0;\, 1 \leq x \leq n. \tag{31}$$

In Eq. (28), when $k_1(i, R_n)$ is at its maximum $R_N$, decreasing the rate of the cognitive link has no effect on the rate of the primary link and hence $\Delta k_1^{avg} / \Delta k_2^{avg} = 0$.

In brief, the proposed algorithm first assigns the maximum rate $R_N$ to the cognitive link in all regions.

Next, in several steps based on the assigned decision variables to the regions, it attempts to reduce the rate of the cognitive link minimally in order to satisfy the constraints. The steps of the algorithm are detailed below.

*Proposed Algorithm:*

1. Set $k_2(i) = R_N, 1 \leq i \leq V_0$.

2. Compute $k_1(i), 1 \leq i \leq V_0$ based on Eq. (15) and $k_1^{avg}$ and $p_2^{avg}$ based on Eqs. (23) and (26).

3. If none of the constraints is satisfied ($k_1^{avg} \leq E_1$ and $p_2^{avg} \geq P_2^{max}$) go to step 4. Else if only the constraint on average spectral efficiency of link 1 is not satisfied ($k_1^{avg} \leq E_1$ and $p_2^{avg} \leq P_2^{max}$) go to step 8. Else if only the constraint on the average power of the link 2 is not satisfied ($k_1^{avg} \geq E_1$ and $p_2^{avg} \geq P_2^{max}$) go to step 12. Otherwise, ($k_1^{avg} \geq E_1$ and $p_2^{avg} \leq P_2^{max}$) go to step 16.

4. Compute values of $d_3(i), 1 \leq i \leq V_0$ using Eq. (30).

5. Find $i_m = \arg\max_i d_3(i)$. If $d_3(i_m) = 0$, go to step 17 else, if $k_2(i_m) = R_n$, set $k_2(i_m) = R_{n-t}$, where $t$ is given in Eq. (31).

6. Update $k_1^{avg}$ and $p_2^{avg}$ using the next equations.

$$k_1^{avg} \leftarrow k_1^{avg} - (k_1(i, R_n) - k_1(i, R_{n-t})) \, pr(i) \qquad (32)$$

$$p_2^{avg} \leftarrow p_2^{avg} - P_1 (g_{B_2}(R_n) - g_{B_2}(R_{n-t})) \, pr(i) \, p(i) \qquad (33)$$

and $d_3(i_m)$ based on Eq. (30).

7. If $k_1^{avg} \leq E_1$ and $p_2^{avg} \geq P_2^{max}$, go to step 5 else go to step 3.

8. Compute $d_1(i), 1 \leq i \leq V_0$, using Eq. (28).

9. Find $i_m = \arg\max_i d_1(i)$. If $d_1(i_m) = 0$, go to step 17 else, if $k_2(i_m) = R_n$, set $k_2(i_m) = R_{n-t}$, where $t$ is given in Eq. (31).

10. Update $k_1^{avg}$ using Eq. (32) and $d_1(i_m)$ based on Eq. (28).

11. If $k_1^{avg} \leq E_1$ go to step 9, else go to step 16.

12. Compute $d_2(i, 1), 1 \leq i \leq V_0$, based on Eq. (29).

13. Find $i_m = \arg\max_i d_2(i, 1)$. If $d_2(i_m, 1) = 0$, go to step 17 else, if $k_2(i_m) = R_n$, set $k_2(i_m) = R_{n-1}$.

14. Update $p_2^{avg}$ using Eq. (33) and $d_2(i_m)$ using Eq. (29).

15. If $p_2^{avg} \geq P_2^{max}$ go to step 13.

16. The desired $k_2(i)$'s are obtained. End.

17. The constraints cannot be provisioned. End.

*Theorem 1:* The proposed algorithm provides the optimum solution for the optimization problem (27), when the regions are sufficiently small and the next condition is satisfied.

$$k_1(i, k_2(i)) < R_N \implies d_1(i) = D_0 = \text{cte}, \quad 1 \leq i \leq V_0, \forall\, k_2(i) \tag{34}$$

*Proof:* The proof is provided in Appendix C.

### 3) Complexity of the proposed algorithm

In each iteration of the algorithm, a search among $V_0$ numbers and two or three operations to modify $p_2^{avg}$, $k_2^{avg}$ and the corresponding decision variable are performed. Therefore, the complexity is $O(V_0)$ per iteration. The maximum number of iterations is equal to the maximum possible decrements in the values of $k_2(i), 1 \leq i \leq V_0$, from the initial point. The number of iterations is $O(V_0)$, and therefore, the worst case complexity of the algorithm is $O(V_0^2)$. It is noteworthy that the complexity of an exhaustive search to solve this problem is $O(N^{V_0})$.

## IV. PERFORMANCE ANALYSIS OF THE INTERWEAVE APPROACH

For comparison, we consider the interweave approach for the cognitive transmission within the system model described in section II.B. In the interweave approach, the links are not to transmit simultaneously and hence ideally do not interfere with each other. Therefore, for the system model under consideration with a constant power primary radio, interweave transmission in its strict sense leads to zero spectral efficiency for the cognitive link. However, to set an upper bound for the performance of this approach for comparison, we consider an ideal case in which the primary link transmits with its maximum possible rate only in some portions of the spectrum (time or frequency) to satisfy its minimum required spectral efficiency. In turn, the cognitive radio in a fully coordinated manner, i.e., with ideal and instantaneous sensing, transmits in the remaining spectrum holes [11]. In this case, if $\lambda \leq 1$ is the portion of the primary link activity, the average power constraint becomes $P_1/\lambda$ for the primary link and $P_2^{max}/(1-\lambda)$ for the cognitive link during their activity. The average spectral efficiency of the primary link is given by

$$k_1^{avg} = \lambda \left(\sum_{n=1}^{N} R_n \times \text{prob}\{v_{1,n} \leq P_1 s_{11}/(\lambda N_0) < v_{1,n+1}\}\right), \tag{35}$$

in which $v_{1,n} = g_{B_i}(R_n)$ as is computed in (9) and $\lambda$ is computed such that the minimum required average spectral efficiency for the primary radio is provided. In the constant power scheme, the average spectral efficiency of the cognitive link is given by

$$k_2^{avg} = (1-\lambda)\left(\sum_{n=1}^{N} R_n \times \text{prob}\{v_{2,n} \leq P_2^{max} s_{22}/((1-\lambda)N_0) < v_{2,n+1}\}\right), \tag{36}$$

where $v_{2,n} = g_{B_2}(R_n)$ as given in (9).

In the adaptive power scheme, the average rate of the cognitive link is

$$k_2^{avg} = (1-\lambda)\left(\sum_{n=1}^{N} R_n \times \text{prob}\{v'_{2,n} \leq s_{22}/N_0 < v'_{2,n+1}\}\right). \tag{37}$$

where $v'_{2,n}$ as computed in [6], is set to maximize the average spectral efficiency of a single link while observing the transmission power and BER constraints.

## V. Performance Evaluation

We use the AMC transmission modes of the IEEE 802.11a standard for the performance evaluation. There are eight modes set up based on different convolutionally coded QAM modulations with rates $R_i \in \{0, 0.5, 0.75, 1, 1.5, 2, 3, 4\}$. The model parameters according to Eq. (4) are derived in [18]. To evaluate the proposed schemes, we consider a system as in Fig. 1 and described in section II, with parameters $\bar{s}_{11} = \bar{s}_{22} = 1$, $\bar{s}_{12} = \bar{s}_{21} = 0.03$, $B_1 = B_2 = 10^{-5}$, $P_1 = 1$ and $P_2^{max} = 2$.

In Fig. 3, the average spectral efficiency of the cognitive link using the proposed interfering scheme with power adaptation is depicted versus the minimum required average spectral efficiency of the primary link. The results are presented for different number of regions ($V_0$). It is clear that choosing $V_0 = 300$, provides accurate results and any further increase leads to only negligible performance improvement. To evaluate the proposed solution for the optimization problem, we also obtained the solution with an exhaustive search algorithm. Comparisons in Fig. 3 indicate that the results obtained by the proposed algorithm for the optimization problem with sufficiently large number of regions are very close to the optimum solution.

In Fig. 4, the performance of the system is depicted for different approaches in the case with adaptive power cognitive radio. It is observed that the proposed interfering transmission scheme provides considerable performance improvement compared to the interweave approach. In Fig. 5, the same comparison is presented for the case with the constant power cognitive radio.

In Fig. 6, the performance of the proposed interfering scheme with power adaptation is depicted for different power constraints on the cognitive radio. As evident, increasing the minimum required average rate for the primary link, reduces the average rate achieved by the cognitive link even when it benefits from a large power constraint. On the other hand, the power constraint on the cognitive link limits the maximum average spectral efficiency of the cognitive link even when the required rate of the primary link is small; this is because of a primary radio, which transmits with a constant power (without turn off). The level at which the rate of the cognitive radio saturates is larger, when its power constraint is less stringent.

In Fig. 7, the case with a large scale path-loss model is considered, i.e., $\bar{s}_{ij} = S_0/d_{ij}^U$, where $d_{ij}$ is the distance between Tx$_i$ and Rx$_j$, U is the path loss exponent (here $U = 3$) and $S_0$ is a constant parameter. The transceivers are positioned on the vertices of a normalized rectangle, i.e., $d_{11} = d_{22} = 1$ and $d_{12} = d_{21} = \sqrt{1 + d^2}$, where $d$ is the distance between the transmitters (receivers). As expected, since the interfering signal is treated as noise, increasing $d$ reduces the interference level and hence improves the performance.

The performance improvement provided by the proposed interfering transmission scheme compared to the interweave transmission is in accord with what is anticipated in information theory for an interference channel. Specifically, for a weak interference channel with constant gains, it is shown that concurrent transmission and treating interference as noise results in higher rates compared to a TDMA/FDMA scheme [4]. It is noteworthy that, finding the spectrum holes in the interweave approach still needs a computationally complex spectrum sensing block, which is not required in the proposed schemes.

## VI. Conclusions

For a wireless fading network with a primary and a cognitive radio, a cognitive transmission scheme based on interfering transmission was developed. The cognitive radio transmits with the primary user simultaneously in the same frequency band. Using adaptive rate and power transmission, the average spectral efficiency of the cognitive link was maximized in a way that a minimum required average spectral efficiency for the primary link is provided and the power and BER constraints on both links are observed.

For the future works, we consider an adaptive power primary radio and provisioning QoS constraints for the users. Other fruitful research directions include devising effective and practical schemes for SNIR feedback and addressing considerations for employing the proposed interfering cognitive radio transmission scheme in a network with large number of users.

## APPENDIX A: COMPUTING PDF OF A RATIONAL FUNCTION OF EXPONENTIAL RANDOM VARIABLES

*Lemma 1:* If $x_i, i = 1,2$ is an exponentially distributed random variable with mean $\bar{x}_i$, i.e., $f_{x_i}(x_i) = \frac{1}{\bar{x}_i}\exp\left(-\frac{x_i}{\bar{x}_i}\right)$ and $y = Q_1 x_1/(Q_2 x_2 + Q_3)$, where $Q_1, Q_2$ and $Q_3$ are constants, the PDF of $y$ is given by:

$$f_y(y_0) = \left(\frac{1}{\bar{x}_1}\frac{1}{\bar{x}_2}Q_3/(\frac{1}{\bar{x}_2}Q_1 + \frac{1}{\bar{x}_1}y_0 Q_2) + \frac{Q_2}{Q_1}\frac{1}{\bar{x}_2}\frac{1}{\bar{x}_1}/\left(\frac{1}{\bar{x}_2} + \frac{1}{\bar{x}_1}\frac{y_0 Q_2}{Q_1}\right)^2\right) \times \exp\left(-\frac{1}{\bar{x}_1}\frac{y_0 Q_3}{Q_1}\right).$$

*Proof:* Considering the cumulative density function (CDF) of $y$, we have:

$$\text{prob}\{y \le y_0\} = \text{prob}\left\{x_1 \le \frac{y_0(Q_2 x_2 + Q_3)}{Q_1}\right\} = \int_0^\infty \int_0^{\frac{y_0(Q_2 x_2 + Q_3)}{Q_1}} f_{x_1}(x_1) f_{x_2}(x_2)\, dx_1\, dx_2 = \int_0^\infty \left(1 - \exp\left(-\frac{1}{\bar{x}_1} \times \frac{y_0(Q_2 x_2 + Q_3)}{Q_1}\right)\right) f_{x_2}(x_2)\, dx_2 = 1 - \exp\left(-\frac{1}{\bar{x}_1}\frac{y_0 Q_3}{Q_1}\right) \times \frac{1}{\bar{x}_2}/(\frac{1}{\bar{x}_2} + \frac{1}{\bar{x}_1}\frac{y_0 Q_2}{Q_1}).$$

Differentiating this CDF w.r.t. $y_0$, the PDF function of $y$ is obtained. ∎

## APPENDIX B: COMPUTATION OF ASSIGNED VARIABLES TO THE REGIONS IN III-B-2

In this section, the parameters $pr(i), p(i)$ and $k_1(i, k_2(i) = 0)$ are derived in simplified forms. Also, to compute $k_1(i, k_2(i) = 0)$, $f_{\alpha,\beta,s_{11}}(\alpha, \beta, s_{11})$ is obtained.

To compute $pr(i)$ and $p(i)$ for a region that is determined by border lines $\beta/\alpha = W_j$ and $\frac{\beta}{\alpha} = W_{j+1}, 0 \le j \le L$ and curves $\alpha\beta = T_{h-1}$ and $\alpha\beta = T_h$ ($\{T_h, 0 \le h \le C, T_0 \le \cdots \le T_C\} = \{Z_a, 0 \le a \le M\} \cup \{Z'_b, 1 \le b \le C - M\}$, see Eqs. (19), (20) and (21)), we change the integration variables as $x^2 = \alpha\beta; y^2 = \beta/\alpha$ and obtain:

$$pr(i = jL + h) = \iint_{Reg(i)} f_\alpha(\alpha) f_\beta(\beta)\, d\alpha\, d\beta = \int_{\sqrt{W_j}}^{\sqrt{W_{j+1}}} \int_{\sqrt{T_h}}^{\sqrt{T_{h+1}}} 2ABxy/((Ay + x)^2 (B + xy)^2)\, dx\, dy,$$

$$p(i = jL + h) = \iint_{Reg(i)} 1/\beta \times f_\alpha(\alpha) f_\beta(\beta)\, d\alpha\, d\beta = \int_{\sqrt{W_j}}^{\sqrt{W_{j+1}}} \int_{\sqrt{T_h}}^{\sqrt{T_{h+1}}} 2AB/((Ay + x)^2 (B + xy)^2)\, dx\, dy.$$

where $A = \bar{s}_{11}/\bar{s}_{21}$, $B = \bar{s}_{22}/\bar{s}_{12}$ and $1 \le i \le V_0 = LC$. The resulting equations for $pr(i)$ and $p(i)$ can be simply evaluated by numerical integration methods.

To compute $k_1(i, k_2(i) = 0)$, as in (24), the next probabilities should be evaluated using $f_{\alpha,\beta,s_{11}}(\alpha, \beta, s_{11})$.

$$\text{prob}\{v_{1,n} \le P_1 s_{11}/N_0 < v_{1,n+1} | \alpha, \beta \in Reg(i)\} = \text{prob}\left\{v_{1,n} \le \frac{P_1 s_{11}}{N_0} < v_{1,n+1}, \alpha, \beta \in Reg(i)\right\}/pr(i).$$

We have $f_{\alpha,\beta,s_{11}}(\alpha,\beta,s_{11}) = f_\beta(\beta) \times f_{\alpha,s_{11}}(\alpha,s_{11})$, as $\beta$ is independent of $\alpha$ and $s_{11}$ and we have:

$$f_{\alpha,s_{11}}(a_0,s_0) = \frac{\partial^2}{\partial a_0 \partial s_0} \text{prob}\{\alpha < a_0, s_{11} < s_0\} = \frac{\partial^2}{\partial a_0 \partial s_0} \int_0^{s_0} \int_{\frac{P_1 s_{11}}{a_0}}^{\infty} \frac{1}{\bar{s}_{11}} \frac{1}{\bar{s}_{21}} \exp\left(-\frac{s_{11}}{\bar{s}_{11}}\right) \exp\left(-\frac{s_{21}}{\bar{s}_{21}}\right) ds_{21} ds_{11} =$$

$$\frac{1}{\bar{s}_{11} \times \bar{s}_{21}} \frac{P_1 s_0}{a_0^2} \exp\left(-s_0 \left(\frac{1}{\bar{s}_{11}} + \frac{1}{\bar{s}_{21}} \frac{P_1}{a_0}\right)\right).$$

Using joint PDF of $\alpha, \beta$ and $s_{11}$ and by changing integration variables as $x^2 = \alpha\beta; y^2 = \beta/\alpha$, we have:

$$\text{prob}\{v_{1,n} \leq P_1 s_{11}/N_0 < v_{1,n+1}, \alpha, \beta \in Reg(i)\} = \iint_{Reg(i)} \int_{v_{1,n} N_0/P_1}^{v_{1,n+1} N_0/P_1} f_{\alpha,s_{11}}(a,s_{11}) f_\beta(\beta) \, d\alpha \, d\beta ds_{11}$$

$$= \int_{\sqrt{W_j}}^{\sqrt{W_{j+1}}} \int_{\sqrt{T_h}}^{\sqrt{T_{h+1}}} \int_{v_{1,n} N_0/P_1}^{v_{1,n+1} N_0/P_1} P_1 s_{11} \frac{2y/x}{\bar{s}_{11} \times \bar{s}_{21}} \exp\left(-s_{11}\left(\frac{1}{\bar{s}_{11}} + \frac{1}{\bar{s}_{21}} \frac{P_1 y}{x}\right)\right) \frac{B}{(B+xy)^2} dx dy ds_{11}.$$

The above equation can be simply evaluated by numerical integration methods.

### APPENDIX C: PROOF OF THEOREM 1

This section provides a proof to theorem 1. To this end, we first consider two lemmas:

*Lemma 2:* Suppose that $\mathcal{A}$ is an ordered set of fractions $\left\{\frac{a_i}{b_i}\right\}, 1 \leq i \leq N$, such that $a_i > 0, b_i > 0$ for $1 \leq i \leq N$ and $\frac{a_N}{b_N} \leq \ldots \leq \frac{a_2}{b_2} \leq \frac{a_1}{b_1}$. We have $\frac{a_N}{b_N} \leq \frac{\sum_{i=1}^N a_i}{\sum_{i=1}^N b_i} \leq \frac{a_1}{b_1}$.

*Proof:*

If $\frac{a_1}{b_1} \geq \frac{a_2}{b_2}$ and $a_i, b_i > 0$ for $1 \leq i \leq 2$, then $\frac{a_1}{b_1} \geq \frac{a_1+a_2}{b_1+b_2} \geq \frac{a_2}{b_2}$ (proof is straight forward by comparing the multiplication of the mean terms and the extern terms of fractions). To prove the lemma by induction we also demonstrate that if the above fact is true for K fractions, it is also true for $K+1$ fractions, as follows:

$$\frac{a_{K+1}}{b_{K+1}} \leq \frac{a_K}{b_K} \leq \ldots \leq \frac{a_1}{b_1} \text{ and } \frac{a_K}{b_K} \leq \frac{\sum_{i=1}^K a_i}{\sum_{i=1}^K b_i} \leq \frac{a_1}{b_1} \Rightarrow \frac{a_{K+1}}{b_{K+1}} \leq \frac{a_K}{b_K} \leq \frac{\sum_{i=1}^K a_i}{\sum_{i=1}^K b_i} \leq \frac{a_1}{b_1} \xrightarrow{\text{lemma 2 (N=2)}} \frac{a_{K+1}}{b_{K+1}} \leq \frac{\sum_{i=1}^{K+1} a_i}{\sum_{i=1}^{K+1} b_i} \leq \frac{a_1}{b_1}.$$

The proof is complete. ∎

*Lemma 3:* Consider $\mathcal{A}$ as defined in lemma 2, $\mathcal{B}$ a subset of $\mathcal{A}$ containing its L largest elements, $\mathcal{B} = \left\{\frac{a_1}{b_1}, \frac{a_2}{b_2}, \ldots, \frac{a_L}{b_L}\right\}$ and $\mathcal{C}$ a subset of $\mathcal{A}$ with M elements, $\mathcal{C} = \left\{\frac{a'_1}{b'_1}, \frac{a'_2}{b'_2}, \ldots, \frac{a'_M}{b'_M}\right\}$ such that $\mathcal{B} \cap \mathcal{C} \neq \mathcal{B}, \mathcal{C}$.

If $\sum_{i=1}^M b'_i = \sum_{i=1}^L b_i$ then $\sum_{i=1}^M a'_i \leq \sum_{i=1}^L a_i$ and if $\sum_{i=1}^M a'_i = \sum_{i=1}^L a_i$ then $\sum_{i=1}^M b'_i \geq \sum_{i=1}^L b_i$.

*Proof:*

Assume $\mathcal{B} - \mathcal{C} = \left\{\frac{e_1}{d_1}, \ldots, \frac{e_{N_1}}{d_{N_1}}\right\}$ and $\mathcal{C} - \mathcal{B} = \left\{\frac{e'_1}{d'_1}, \ldots, \frac{e'_{N_2}}{d'_{N_2}}\right\}$ and $\mathcal{B} \cap \mathcal{C} = \left\{\frac{e''_1}{d''_1}, \ldots, \frac{e''_{N_3}}{d''_{N_3}}\right\}$. We have:

$$\frac{\sum_{i=1}^{N_2} e'_i}{\sum_{i=1}^{N_2} d'_i} \leq \max\{\mathcal{C} - \mathcal{B}\} \leq \min\{\mathcal{B} - \mathcal{C}\} \leq \frac{\sum_{i=1}^{N_1} e_i}{\sum_{i=1}^{N_1} d_i}. \tag{38}$$

The first and the last inequalities are directly obtained from lemma 2. The second inequality is proved as follows. If a member of $\mathcal{A}$ is larger than one of the members of $\mathcal{B}$, it is also a member of $\mathcal{B}$, as $\mathcal{B}$ comprises the L largest members of $\mathcal{A}$. Due to this fact and because $\max\{\mathcal{C} - \mathcal{B}\} \in \mathcal{A}$ and $\min\{\mathcal{B} - \mathcal{C}\} \in \mathcal{B}$, $\max\{\mathcal{C} - \mathcal{B}\} > \min\{\mathcal{B} - \mathcal{C}\}$ results in $\max\{\mathcal{C} - \mathcal{B}\} \in \mathcal{B}$. This is a contradiction and therefore the second inequality is true.

If $\sum_{i=1}^{M} b'_i = \sum_{i=1}^{N} b_i$, then $\sum_{i=1}^{N_1} d_i = \sum_{i=1}^{N_2} d'_i$ and due to (38), we have $\sum_{i=1}^{N_1} e_i \geq \sum_{i=1}^{N_2} e'_i$; Adding $\sum_{i=1}^{N_3} e''_i$ to both sides results in $\sum_{i=1}^{M} a'_i \leq \sum_{i=1}^{N} a_i$. In a similar way, the second part of the lemma is proved. ∎

*Proof of theorem 1:*

In the proposed algorithm, in the first step, the rate of the cognitive link in all regions is set to maximum, i.e., $R_N$. Subsequently, if both spectral efficiency and power constraints in (27) are satisfied, the maximum achievable average spectral efficiency for the cognitive link is obtained. It is clear that in all cases to increase $k_1^{avg}$ or decrease $p_2^{avg}$, $k_2^{avg}$ is to be reduced. This implies that the values of $k_2(.)$ are to be decreased in certain regions. The proposed algorithm consists of three parts: part 1 (steps 4-7), when both constraints are not satisfied; part 2 (steps 8-11), when only the power constraint of the cognitive link is provisioned; part 3 (steps 12-15), when only the average spectral efficiency of the primary link is satisfied. The process of satisfying both constraints can be considered as taking a number of basic changes as defined below.

*Definition: Basic Changes*

A basic change is a minimum possible decrease in the assigned rate to the cognitive link in a region, such that it results in an increase in the rate of the primary link or a decrease in the power of the cognitive link. Therefore, there are two types of basic changes; if the cognitive link rate in $Reg(i)$ is $k_2(i) = R_n > 0$, then a power effective basic change is to set $k_2(i) = R_{n-1}$ and a rate effective basic change is to set $k_2(i) = R_{n-t}$, while $t$ is defined in Eq. (31). The algorithm as presented in section III.B.3 identifies a specific feasible basic change in each iteration.

Note that making basic changes in a region is ordered, i.e., basic changes should be done step by step from a higher rate to a lower rate. For example, if $k_2(i) = R_n > 0$, for each $b, c > 0$ a basic change to decrease $k_2(i)$ from $R_{n-b}$ to $R_{n-b-c}$ is feasible, when the basic changes to reduce $k_2(i)$ from $R_n$ to $R_{n-b}$ are taken, previously.

To collect all rate effective basic changes, for each region we start with $k_2(i) = R_N$ and reduce it step by step (with a proper $t$ defined in Eq. (31)) untill $k_2(i) = 0$.

To prove the algorithm, we start with the case in which the cognitive transmitter power constraint is not satisfied and the minimum required average spectral efficiency for the primary link is provided (part 3). The goal is to reduce $p_2^{avg}$ to $P_2^{max}$, in a way that the reduction in $k_2^{avg}$ is minimized. Suppose that the set of all power effective basic changes are sorted in descending order based on their corresponding $d_2(i, 1) = \Delta p_2^{avg}/\Delta k_2^{avg}$ (Eq. (28)). For the desired set of power effective basic changes, the summation of the nominators of $d_2(i, 1)$ is to be $p_2^{avg} - P_2^{max}$ (this is possible as the regions are sufficiently small), while the summation of the denominators (decrease in $k_2^{avg}$) is minimum. Noting lemma 3, if the selected changes are the ones with the largest $d_2(i, 1)$, we have the minimum decrease in $k_2^{avg}$. The proposed algorithm makes such a selection. To complete the proof, we should show that the proposed set of basic changes is feasible based on the ordering condition, i.e., the next inequality is valid:

$$\frac{\left(g_{B_2}(R_n)-g_{B_2}(R_{n-b})\right)}{R_n-R_{n-b}} \geq \frac{\left(g_{B_2}(R_{n-b})-g_{B_2}(R_{n-b-c})\right)}{R_{n-b}-R_{n-b-c}}, \forall b, c > 0. \tag{39}$$

The above fact, for $b = c = 1$ can be numerically validated. Hence, the correctness of the inequality is proven using lemma 2 for all $b, c$. Based on (39), $d_2(i, b)$ when $k_2(i) = R_n$ is higher than $d_2(i, c)$ when $k_2(i) = R_{n-b}$. This affirms that the power effective basic changes are only taken in order.

For the case in which the minimum required average spectral efficiency for the primary link is not provided and the power constraint on the cognitive transmitter is satisfied, if the average spectral efficiency of the primary link at the beginning of the algorithm is $E_1^0$, the rate of the cognitive link is to decrease more than $(E_1^0 - E_1)/D_0$ in order to increase the average spectral efficiency of the primary link to $E_1$. This is due to the condition in (34). Therefore, the minimum decrease in the cognitive link rate occurs when the rate effective basic changes with $d_1(i) > 0$ are selected. The proposed algorithm suggests such a selection.

In the case that none of the constraints are satisfied, in a similar way to the previous part, we can show that the rate of the cognitive link is to decrease more than $(E_1^0 - E_1)/D_0$. Therefore, the minimum decrease in the cognitive link rate occurs when rate effective basic changes are selected, a basic change that does not affect $k_1^{avg}$ ($d_1(.) = 0$), results in an extra reduction in the $k_2^{avg}$ compared to $(E_1^0 - E_1)/D_0$. The algorithm in part 1 proceeds to satisfy at least one of the constraints. In this part of the algorithm, among the entire rate effective basic changes, the ones with the largest $d_3(.)$ are selected. Suppose that the power constraint is met first during the iterations in part 1, the algorithm proceeds in part 2 and because all of the selected changes in part 1 and 2 have $d_1(.) = D_0$, the minimum reduction in the cognitive link rate, $(E_1^0 - E_1)/D_0$, occurs and therefore the algorithm is optimum.

We also consider the case in which the rate constraint is met first after a number of iterations in part 1. In this case, because the basic changes with the largest $d_2(.)$ are selected, based on lemma 3, we have the maximum possible decrease in the $p_2^{avg}$ compared to any other set of changes with the same decrease in $k_2^{avg}$. Feasibility of the proposed basic changes can be shown using (39). In the continuation of the algorithm in part 3, as justified before, the smallest decrease in the $k_2^{avg}$ takes place so that the power constraint is satisfied.

In the algorithm, we may encounter the case where $\max_i d_l(i,.) = 0$ for $l = 1, 2$ or $3$. This indicates that the constraints cannot be satisfied and therefore a feasible solution does not exist. ∎

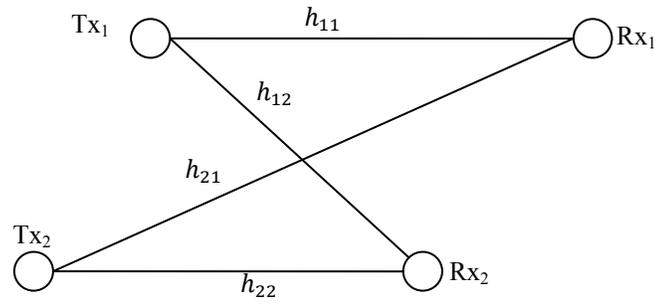

Fig. 1. System configuration: a primary and a cognitive radio

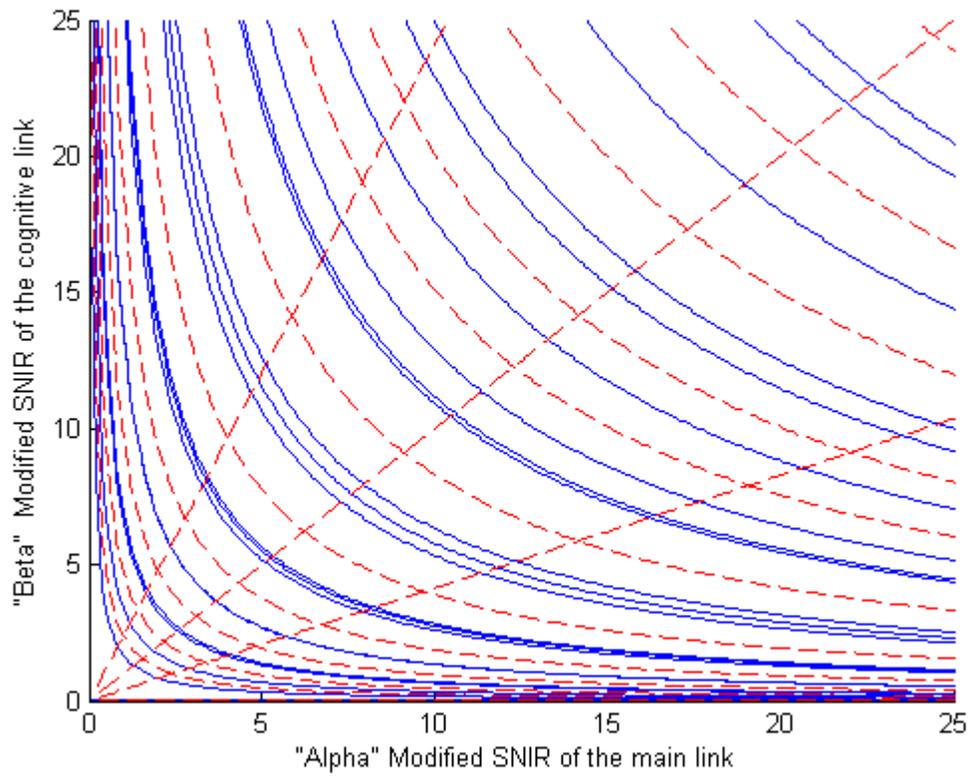

Fig. 2. Partitioning of $\alpha - \beta$ plane into common rate (set) regions.

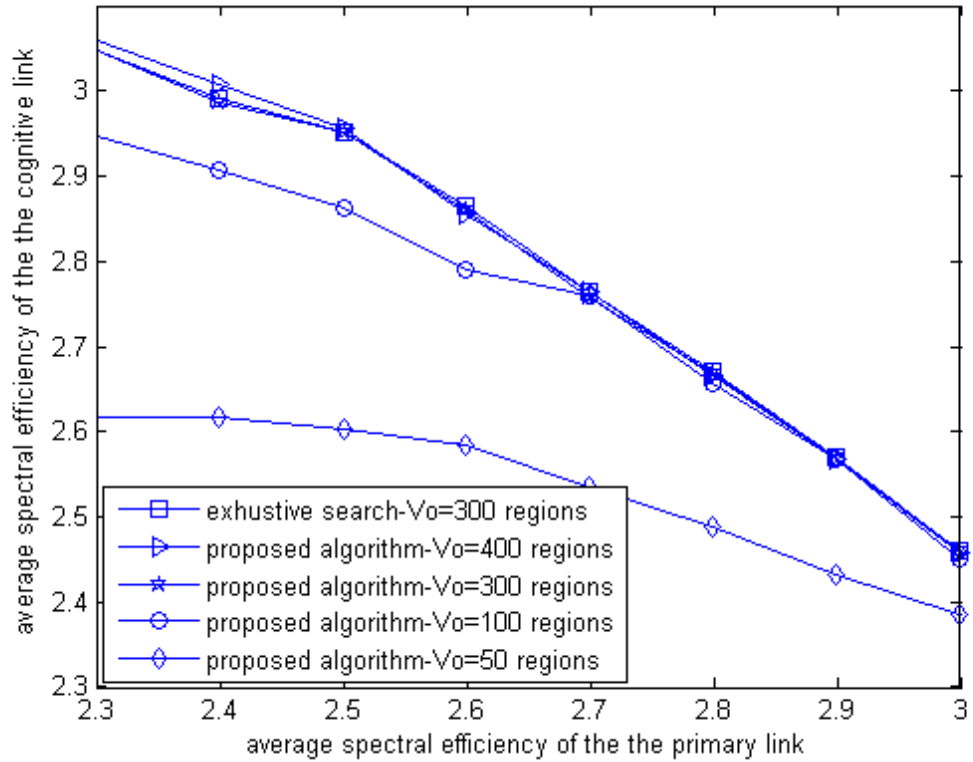

Fig. 3. Spectral efficiency of cognitive vs. primary link; Effect of number of regions and comparison between the proposed algorithm and exhaustive search to solve the optimization problem

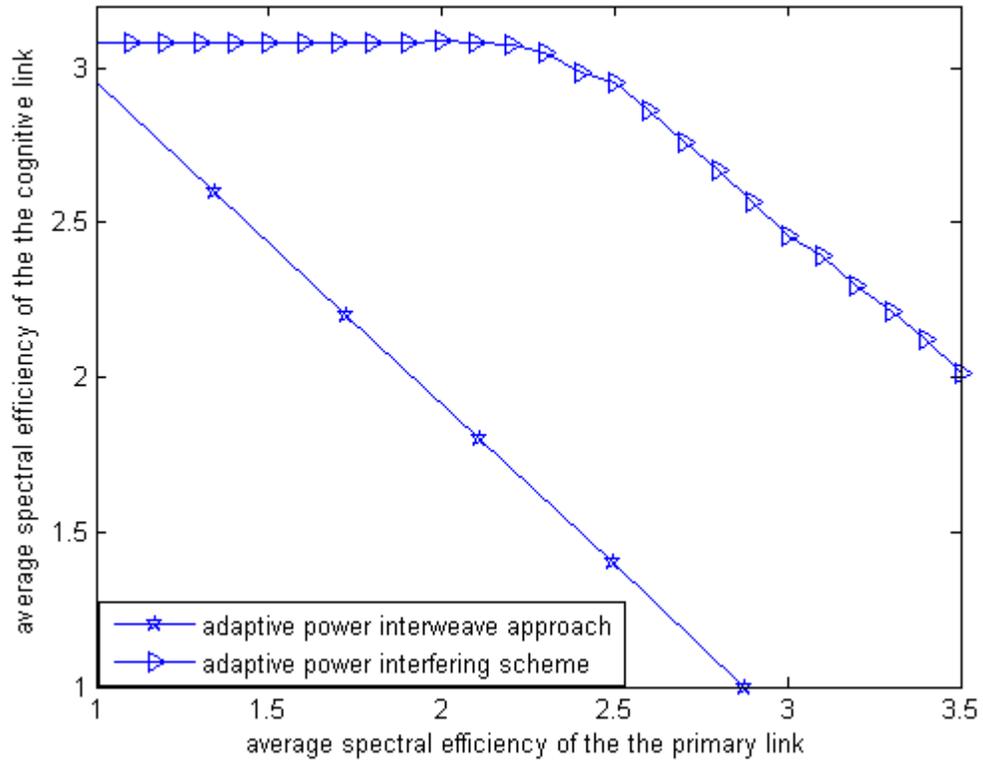

Fig. 4. Performance of the adaptive power interfering cognitive transmission and adaptive power interweave schemes

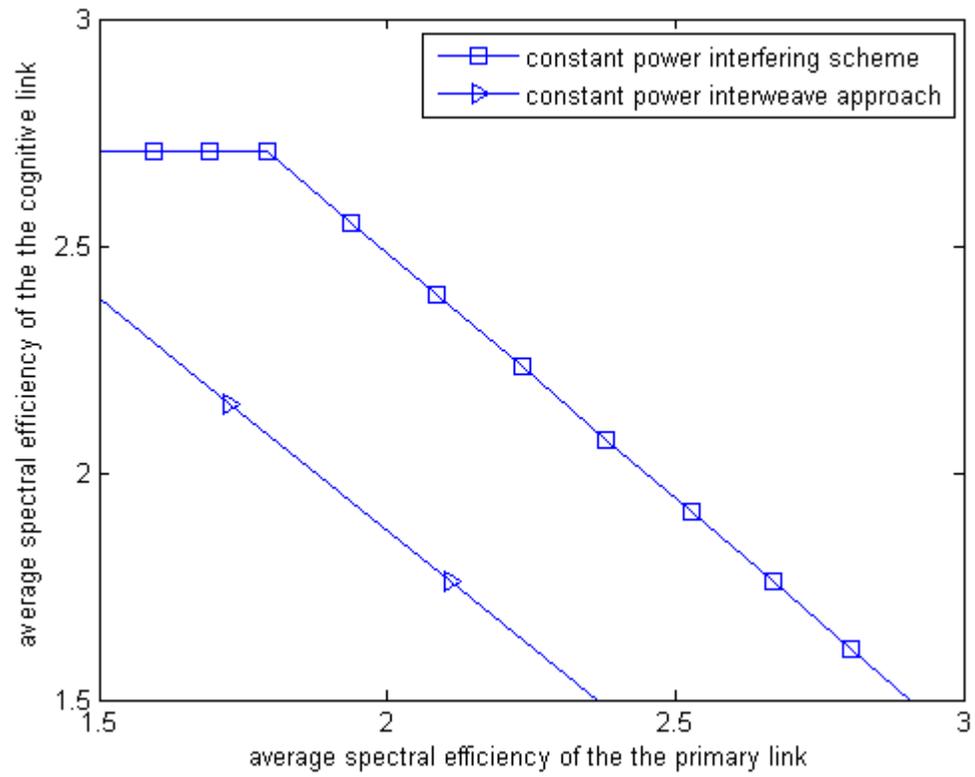

Fig. 5. Performance of the proposed constant power interfering cognitive transmission and constant power interweave schemes

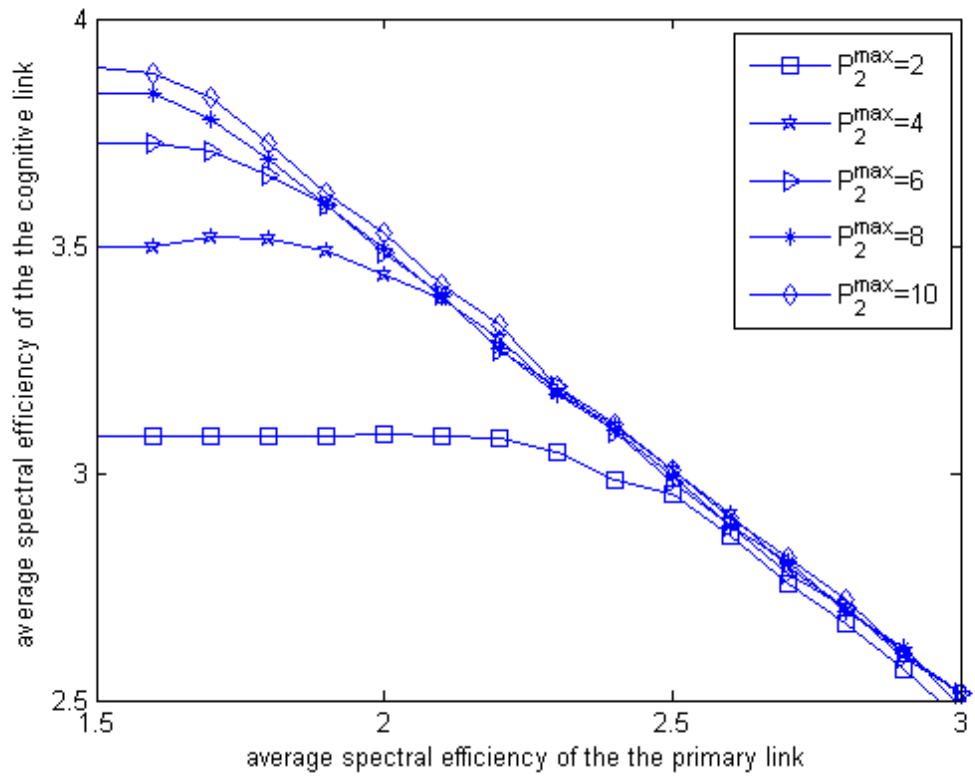

Fig. 6. Performance of the proposed adaptive power interfering cognitive transmission scheme with different power constraints on the cognitive transmitter

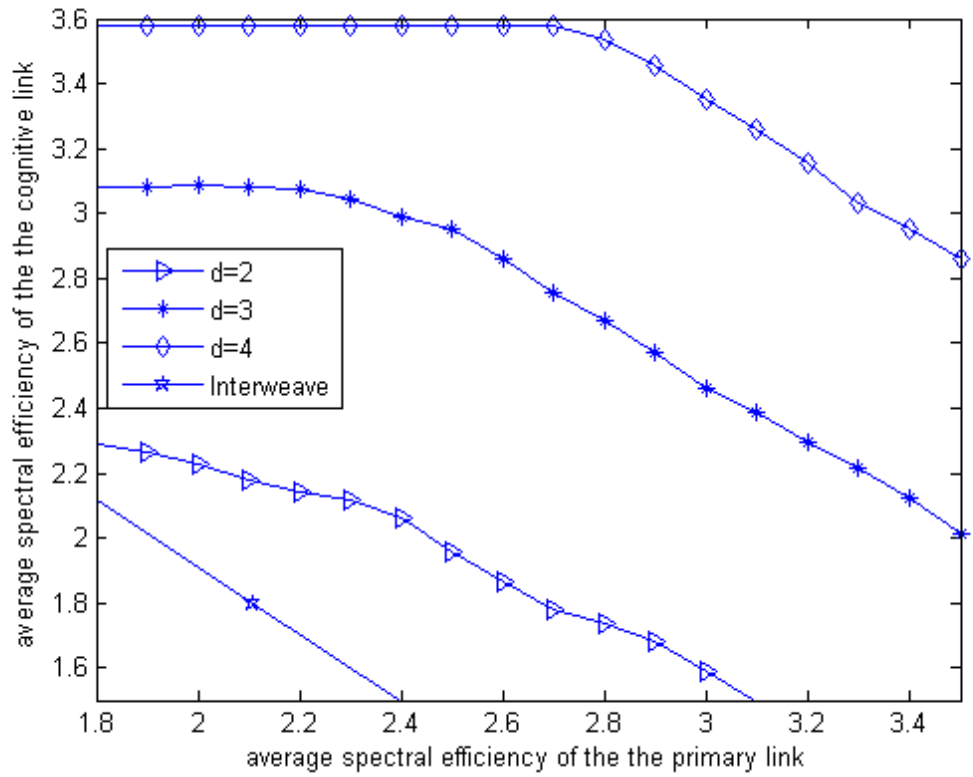

Fig. 7. Spectral efficiency of the cognitive vs. primary links for various *d*'s